\documentclass[amssymb,longbibliography,notitlepage,superscriptaddress,prl,twocolumn,nofootinbib]{revtex4-1}
\usepackage[utf8]{inputenc}
\usepackage{graphicx}
\usepackage{float}
\usepackage{amsmath}

\begin{document}

\title{Reactor neutrino applications and coherent elastic neutrino
  nucleus scattering}

\author{Maitland Bowen}
\email{mebowen@umich.edu}

\affiliation{Center for Neutrino Physics, Virginia Tech, Blacksburg, VA 24061}
\affiliation{University of Michigan, Ann Arbor, MI, 48109}

\author{Patrick Huber}
\email{pahuber@vt.edu}
\affiliation{Center for Neutrino Physics, Virginia Tech, Blacksburg, VA 24061}

\date{\today}

    \begin{abstract}

Potential applications of neutrino detection to nuclear security have
been discussed since the 1970s. Recent years have seen great progress
in detector technologies based on inverse beta decay, with the
demonstration of ton-scale surface-level detectors capable of high
quality neutrino spectrum measurements. At the same time coherent
elastic neutrino nucleus scattering has been experimentally confirmed
in 2017 with neutrinos from stopped pion decay and there is a number
of experiments aimed at seeing this reaction with reactor
neutrinos. The large cross section and threshold-less nature of this
reaction make it plausible to consider it for applications to nuclear
security and here, we present a first direct comparison of the two
reaction modes.

    \end{abstract}
 
\maketitle

\section{Introduction}

Neutrinos were discovered by Cowan and Reines in
1956~\cite{Cowan:1992xc} using neutrinos\footnote{In this letter we
  deal exclusively with electron antineutrinos and for brevity will
  refer to them simply as neutrinos.} from a nuclear reactor and
inverse beta decay (IBD). Nuclear reactors are very bright neutrino
sources with $10^{19}$ neutrinos released per second for 100\,MW of
thermal reactor power. The neutrinos originate in the beta decays of
neutron-rich fission fragments and not in the fission process
itself. As a consequence, the neutrino spectrum and rate are sensitive
to the fissioning isotope via the different fission fragment
yields. In the 1970s, Mikaelyan~\cite{Borovoi:1978} realized that this sensitivity could be exploited to learn about the reactor
state by neutrino observations, giving rise to the field of applied
neutrino physics, for a recent review
see~\cite{Bernstein:2019hix}. One of the major challenges arises from
the fact that detectors for this application have to be able to work
at the Earth's surface and be able to suppress the resulting
backgrounds sufficiently to extract a high-fidelity
signal. Detectors with the requisite characteristics have been
demonstrated only very recently in
2018~\cite{Ashenfelter:2018iov,Haghighat:2018mve}, more than 60 years
after the initial detection of reactor neutrinos.

Coherent elastic neutrino nucleus scattering (CEvNS) was
postulated in 1974~\cite{Freedman:1973yd} and experimentally
confirmed in 2017 by the COHERENT experiment~\cite{Akimov:2017ade}
using neutrinos with 10s of MeV energy. The CEvNS reaction is
interesting for applications because the cross section per unit
detector mass can be two orders of magnitude larger than for IBD,
potentially allowing for detectors in the kilogram range.  This reaction is also threshold-less, potentially providing access to
safeguards-relevant signatures of plutonium
breeding~\cite{Cogswell2016}. The detection of reactor neutrinos, which have
a mean energy of 4\,MeV, via CEvNS has not yet been demonstrated, but
there are several experiments aimed at this
goal~\cite{Agnolet:2016zir,Billard:2016giu,Hakenmuller:2019ecb,Akimov:2019ogx,Aguilar-Arevalo:2019jlr,Angloher:2019flc}.

Physicists in the Soviet Union in 1978 first proposed the use of
neutrinos for remote monitoring of nuclear reactors
\cite{Borovoi:1978}. The typical mixture of isotopes undergoing
fission in a reactor comprises $^{235}$U, $^{238}$U, $^{239}$Pu, and
$^{241}$Pu. The plutonium content is a result of breeding reactions,
which happen in all reactors fueled with natural or low-enriched
uranium and proceeds via neutron captures and beta decays.  Each of
these isotopes also has well-defined and unique neutrino emissions,
and both the energy spectra and the number of neutrinos are
different~\cite{Mueller:2011nm,Huber:2011wv}.  Plutonium-239 has the
lowest average neutrino energy of the four isotopes within the reactor
and this  characteristic  allows for the determination of the plutonium
content of the reactor. The number of neutrinos emitted for each
isotope is also different and the resulting differences in IBD event
rates are shown in Tab.~\ref{tab:IBDrates}. By observing the number of
neutrinos emitted by reactors and the spectra of those neutrinos, the
composition of the reactor fuel and the power level of the reactor can
be determined. As a result, one can deduce whether the reactor could
potentially be producing weapons-grade material, even without access
to records of the reactor history~\cite{Christensen:2013eza}.

Analyses with individual reactors have demonstrated that this method of safeguards
would have provided timely information as to the plutonium production
in the Democratic People's Republic of Korea during the North Korean
nuclear crisis of 1994, even given the limited access inspectors were given
to the reactor \cite{Christensen:2013eza}. Studies have also applied
this method to IR-40, the Iranian heavy water reactor at Arak, and
demonstrated that a neutrino detector can meet or exceed the
verification goals set by the International Atomic Energy Agency
(IAEA) \cite{Christensen:2014pva}.

Potential applications are not limited to reactor power monitoring and
plutonium production monitoring but could include detection of nuclear
waste streams from reprocessing and spent nuclear
fuel~\cite{Christensen:2013eza,Brdar:2016swo}, as well as
long-distance detection, see also
Ref.~\cite{Bernstein:2019hix}. For the two applications considered
here, IBD detectors in mass range of tons are sufficient and detectors
with requisite capabilities have been experimentally demonstrated. At
the same time, the neutrino emission spectra for both reactors and
nuclear wastes have most of their emission below the IBD threshold. In
particular, the neutron capture pathway from uranium to plutonium at a reactor includes beta decays which produce neutrinos of less than
1.2\,MeV. By number, the flux of these ``breeding'' neutrinos exceeds
the flux of fission fragment neutrinos in that energy range
significantly. Therefore, some applications would benefit
significantly if a threshold-less reaction could be
exploited~\cite{Cogswell2016}.

\section{Rates \& spectra}

In IBD an electron antineutrino scatters off a
proton and produces a neutron and positron
\begin{equation}
    \bar\nu_e + p \rightarrow n + e^+  \,.
\end{equation}
Since the neutron is heavier than the proton this reaction has a
threshold energy, which for the proton at rest is given by 
\begin{equation}
    E_\nu^{thr} = \frac{(M_N+m_{e^+})^2 - M_p^2}{2M_p} = 1.806\,\mathrm{MeV}\,.
\end{equation}
Since $m_e,m_\nu\ll m_n,m_p$, the energy of the incoming neutrino
and positron energy have a one-to-one relation
$E_{e^+}=E_\nu-E_\nu^{thr}$. The total cross section at zeroth order
can be expressed as
 \begin{equation}
    \sigma = \frac{2\pi}{m_e^5 f^R\tau_n}E_e p_e \,
\end{equation}
where $\tau_n$ is the measured neutron lifetime and $f^R$=1.7152 is
the phase space factor \cite{Vogel:1999zy}. The cross section for IBD
for reactor neutrinos is approximately $6\times10^{-43}$\,cm$^2$ per
fission. To compute the neutrino event rates and energy spectra from
IBD, we use the cross section from Ref. \cite{Vogel:1999zy} and
antineutrino fluxes for $^{235}$U, $^{238}$U, $^{239}$Pu, and
$^{241}$Pu from a simple summation calculation based on the data in
Ref.~\cite{Huber:2011wv}. The use of this summation calculation allows
us to extend the neutrino spectrum to energies below the IBD threshold,
which is needed for the CEvNS calculations. Note, that this summation
fluxes, as is usual, deviate by about 5\% from the Huber-Mueller
fluxes~\cite{Mueller:2011nm,Huber:2011wv}; however, the relative
properties of the four fissile isotopes are a robust
feature~\cite{Christensen:2013eza}.

We compute the neutrino yield for each isotope through numerical
integration. As a benchmark we chose a 100 MW$_\mathrm{th}$ reactor,
1\,kg CH$_2$ detector, 10\,m standoff from the reactor core and a data
taking period of one year. CH$_2$ is a proxy for an actual organic
scintillator but approximates the proton fraction of most
scintillators. Note, that 100\,MW$_\mathrm{th}$ is typical of
plutonium production reactors and thus is a relevant bench mark. 
\begin{table}[h]
    \centering
    \begin{tabular}{c|rrrr}
        Isotope & $^{239}$Pu & $^{241}$Pu & $^{235}$U & $^{238}$U \\
        \hline
        Events & 288&398&418&636\\
    \end{tabular}
    \caption{IBD event number per kg of CH$_2$ per year at a 100\,MW$_\mathrm{th}$ reactor and at a distance of 10\,m.}
    \label{tab:IBDrates}
\end{table}
The results are shown in Tab.~\ref{tab:IBDrates}, the number of
neutrinos detected with IBD from each isotope in a reactor differs
significantly. At approximately 300 events with our given parameters,
$^{239}$Pu produces the fewest neutrinos, at only about two-thirds the
rate of $^{238}$U. $^{239}$Pu also has the lowest mean energy;
$^{238}$U has the highest mean energy. With observations of both the
event number and the spectra of the neutrinos emitted from a nuclear
reactor, one can deduce both the composition of the mixture within the
reactor and the power level of the reactor itself.

Coherent elastic neutrino-nucleus scattering (CEvNS) was
postulated soon after neutral currents were discovered; it occurs
between a neutrino of any flavor and a target nucleus
\cite{Freedman:1973yd}
\begin{equation*}
    \bar\nu + X \rightarrow \bar\nu + X \,,
\end{equation*}
where the signature is the recoil of the target nucleus $X$. The cross section is approximately given by
\begin{equation}
  \label{eq:cevns}
\frac{d\sigma}{dT}(E_\nu) = \frac{G_F^2}{4\pi}N^2 M \left(1 - \frac{M T}{2E_\nu^2} \right)\,,
\end{equation}
where we have neglected any nuclear form factors; for reactor neutrino
energies this is an excellent approximation. $N$ is the neutron
number, $M$ is the nuclear mass and $T$ the nuclear recoil
energy. CEvNS holds promise for low-energy neutrinos detection due
to the $N^2$ dependence. Despite its high cross section, CEvNS
evaded detection for decades because of the difficulty in detecting
very low nuclear recoil energies. In 2017, the COHERENT collaboration
used a 14.6\,kg CsI[Na] scintillator detector to observe CEvNS for
the first time from neutrinos at the Spallation Neutron Source at Oak
Ridge National Laboratory \cite{Akimov:2017ade}. The relation between
the observable recoil energy $T$ and neutrino energy $E_\nu$ is not
one-to-one and is more similar to the case of Compton-scattering,
therefore the neutrino energy spectrum information is less direct than
in the IBD case.

For CEvNS we use the cross section in Eq.~\ref{eq:cevns} and the
same reactor parameters and standoff as for IBD.  The relevant
observable is the nuclear recoil energy $T$ and for a given neutrino
energy there is a kinematic limit to what the maximum recoil energy
can be
\begin{equation}
  \label{eq:tmax}
T_\mathrm{max} = \frac{E_\nu}{1 + \frac{M_N}{2E_\nu}}\,.
\end{equation}
Unlike IBD, CEvNS can occur on any target and thus there is a wide
range of potential detector materials. We show results for a
selection of commonly used target materials. The cross section peaks at low values of $T$ and therefore, the total event rate sensitively depends on the
low-energy detection threshold for nuclear recoils,
$T_\mathrm{min}$. One common way to quote CEvNS cross sections is
per target nucleus but for practical application the actual target
mass is more critical. Table~\ref{tab:CEvNSrates} shows the neutrino
event number above a given nuclear recoil energy threshold for various
target materials that could serve in detectors for CEvNS. Note,
that we sum the contributions of each stable isotope of each element
weighted by its natural abundance.
\begin{table}[H]
    \centering
    \begin{tabular}{c|r|r|r|r|r|r|r}
            Threshold $[\mathrm{eV}]$ &C&Ne&Si&Ar&Ge &Xe &W\\
        \hline
0&1255 & 2147 & 2958 & 5048 & 9526 & 19033 & 27406\\
10&1223 & 2058 & 2794 & 4669 & 8343 & 15270 & 20462\\
100&1023 & 1565 & 1954 & 2908 & 3913 & 4623 & 4350 \\
1000&335 & 296 & 227 & 169 & 35 & 1 & 0 \\   
    \end{tabular}
    \caption{CEvNS event number per kg per year for fission of
      $^{235}$U at a 100\,MW$_\mathrm{th}$ reactor and at a distance
      of 10\,m as a function of isotope and recoil energy
      threshold. }
    \label{tab:CEvNSrates}
\end{table}

Table \ref{tab:CEvNS_boundary} shows the recoil energies at which a
CEvNS detector with the given isotope as a target will detect the
same number of neutrinos as an IBD detector of the same mass. This is
again using the parameters outlined previously with respect to reactor
power, observation time, distance, and target mass for $^{235}$U. For
example, a germanium detector
will detect fewer neutrinos than a currently operating IBD detector if
the CEvNS detector is unable to observe recoil energies below the
required 496\,eV. Table \ref{tab:CEvNS_boundary} also demonstrates the
linearity of the threshold energy versus the mass number of the
target. Note that we are specifically quoting the nuclear recoil
energy, the electron-equivalent measured energy typically is much
lower due to quenching.

This allows us to make a first observation: almost every potential
target isotope we considered will detect fewer neutrinos than a
currently operating IBD detector if the CEvNS detector is unable
to observe recoil energies below 1\,keV. To put this in
context, the original COHERENT observation was achieved with a recoil
threshold of approximately 5\,keV. It is important to point out that
what matters here is the nuclear recoil threshold where a reactor
CEvNS event can be identified at a reasonable level of background
\emph{and} with good efficiency. Consider for example the recent
CONNIE result~\cite{Aguilar-Arevalo:2019zme}, where the detector has a
threshold of 64\,eV, but only at around 200\,eV the median efficiency
is reached and once quenching is taken into account, this corresponds
to $\sim 1\,\mathrm{keV}$ recoil threshold.
\begin{table}[h]
    \centering
    \begin{tabular}{c|rrrrrrrrr}
        Isotope & C&Ne&Si&Ar&Ge &Xe &W\\
        \hline
        $T_{min}$ [eV] & 791&782&707&677&496&352&281\\
    \end{tabular}
    \caption{The recoil energy threshold at which IBD and the
      CEvNS detection result in the same neutrino event number.}
    \label{tab:CEvNS_boundary}
\end{table}

\begin{figure}[t]
\includegraphics[width=\columnwidth]{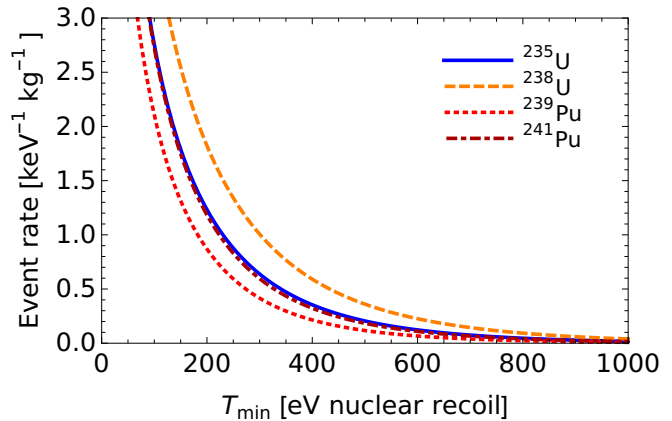}
\caption{\label{fig:spectrum} Shown is the nuclear recoil spectrum on
  germanium in arbitrary units for neutrinos stemming from fission of
  $^{235}$U, $^{239}$Pu, $^{238}$U, and $^{241}$Pu, respectively.}
\end{figure}

Unlike IBD, CEvNS does not have a one-to-one correspondence
between the energy deposited in the detector and the true neutrino
energy. For applications, the potential to track plutonium
production is entirely due to the different energy spectra of the
different fissile isotopes. In Fig.~\ref{fig:spectrum} we show the
recoil energy spectrum in germanium for the usual fissile
isotopes. The spectral differences persist also in the recoil spectrum
and are most prominent in the 100-200\,eV region.

\section{Application relevance}

Research in IBD applications has come a long way since the original
proposal in 1978 and we take the results of Ref.~\cite{Carr:2018tak},
which in turn is based on the performance of the PROSPECT
detector~\cite{Ashenfelter:2018iov}, as our benchmark in
reference to two scenarios relevant to nuclear security:
\begin{itemize}
  \item Reactor power: How long does it take to observe a transition
    from reactor on to off or vice versa? With IBD, this can be achieved with
    1--2 tons of active detector mass over a period of hours to days.
    \item Plutonium production: How long does it take to distinguish a
      new, plutonium-free, reactor core from an old, plutonium-rich,
      core? With IBD, this can be achieved with active detector masses
      of 10--20 tons in a period of weeks to months for most reactor types.
  \end{itemize}
It is of note, that those previous result do include real backgrounds
as measured by PROSPECT and that realistic deployment scenarios have been
considered.

\begin{figure}
\includegraphics[width=\columnwidth]{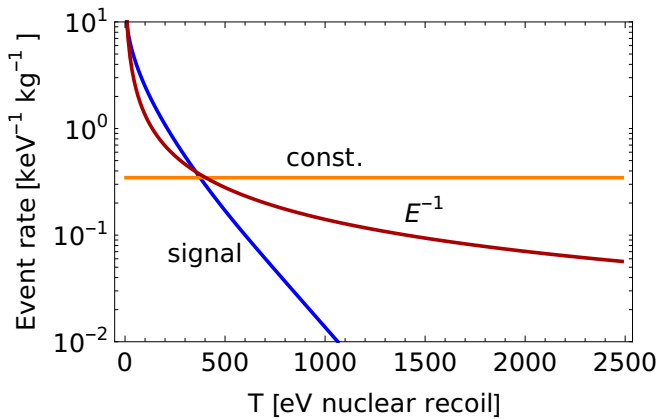}
\caption{\label{fig:bkg} Shown are potential background models for the CEvNS process.}
\end{figure}

CEvNS at reactors has not been yet observed and it is clear that
apart from achieving a low nuclear recoil threshold, mitigation of
backgrounds will be the main challenge, see {\it e.g.}
Ref.~\cite{Hakenmuller:2019ecb}. Apart from the signal to background
(S:B) ratio also the shape of the background matters, especially for
$S:B < 1$. In Fig.~\ref{fig:bkg} we show the signal shape in
comparison to a number of plausible background shapes. In reality the
background will be a mix of shapes since several independent processes
will contribute. Clearly, a $E^{-1}$ background is the most
pernicious, yet seems to be similar to what many experiments see at
very low recoil thresholds, see Ref.~\cite{Kurinsky:2020dpb} for a
recent compilation of low-energy backgrounds.

Having fixed the shape of our background model the only remaining free
parameters are S:B and the recoil energy threshold. We use the same
analysis framework as Ref.~\cite{Carr:2018tak} and can now evaluate
what the ratio of the measurement precision for either reactor power
or plutonium content is for a fixed mass of IBD and CEvNS
detectors. The employed likelihood function is quadratic in the
parameters of interest and hence all precision scales as the inverse
square of detector mass. Therefore, the inverse square of the ratio of
precision for a given measurement expresses how many times smaller or
larger a CEvNS detector would need to be to match the capabilities
of an IBD detector; we call this quantity the mass advantage. If we
neglect backgrounds, we find that the mass advantage becomes unity for
about one half the recoil energy threshold values listed in
Tab.~\ref{tab:CEvNS_boundary}.

This result applies for both power and
plutonium content measurements. This arises from less pronounced
spectrum differences between fissile isotopes in CEvNS. The reason
this also affects the power measurements is the
degeneracy between reactor power and plutonium
content~\cite{Christensen:2013eza}. In the ideal background-free case
the mass advantage for xenon can reach 40 for plutonium-content
determination and up to 80 for power monitoring, albeit at a recoil
threshold of 5\,eV. At the other end of atomic masses, the
mass advantage for silicon is 3--6 at best.

\begin{figure}
\includegraphics[width=\columnwidth]{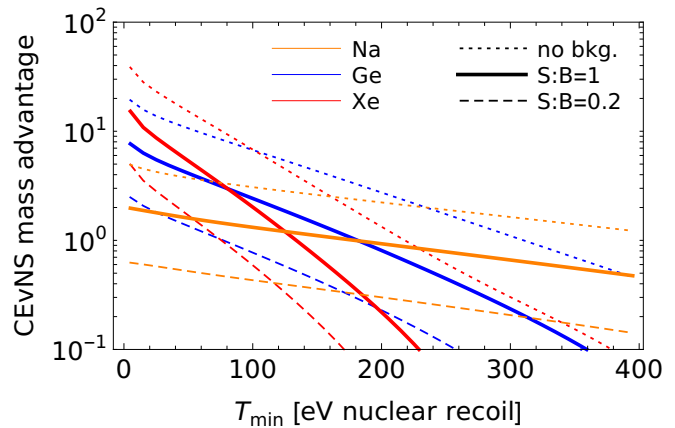}
\caption{\label{fig:massadvan} Shown is the ratio of detector mass
  between CEvNS and IBD to achieve the same precision as the
  corresponding IBD measurement of the reactor plutonium content,
  including a $1/E$ background component with the indicated S:B
  ratio. }
\end{figure}

Taking into account background, we obtain the results shown in
Fig~\ref{fig:massadvan}, which are computed for sodium, germanium, and
xenon for a plutonium content measurement. For S:B=1, which is
essentially what has been demonstrated for IBD, we find that a mass
advantage of unity is reached for a recoil threshold of about
140--300\,eV. The mass advantage climbs to a value of 8--12 for a 5\,eV recoil threshold. Should the background be 5 times or more larger than the
signal, it becomes very difficult to obtain any mass advantage.

\section{Summary}

We have presented a first comparison of the event rates, spectra and
resulting reactor monitoring capabilities of inverse beta decay and the recently confirmed CEvNS process. We find that a nuanced picture arises when we consider both the energy threshold for the detection of nuclear recoil and the actual detector mass, instead of the cross section per target nucleus. To achieve detected event rates per unit detector mass on par with IBD, recoil thresholds of 300--800\,eV are necessary. The information contained in the neutrino energy spectrum regarding reactor plutonium content does persist in
CEvNS, albeit at a lower level due to the unknown energy carried
away by the outgoing neutrino. In a direct comparison of the resulting
ability to measure reactor power and the plutonium content, we find
that CEvNS detectors must achieve recoil energy thresholds as low as 200\,eV to be similar to IBD detectors. Taking into account backgrounds, we find
that CEvNS detectors at best offer a mass advantage of one order
of magnitude assuming that eV-scale recoil thresholds are feasible and
a signal to background ratio of about 1 can be achieved. In our
comparison we did not consider the deployed weight of the entire
detector system or the ability to run remotely with no user
intervention for extended periods of time, which, for instance, might present a challenge for cryogenic detectors. Based on those
results, it appears to us that, for the foreseeable future, the best
use case for CEvNS is not to replace IBD detectors but to
complement them, for instance with the direct detection of neutrinos
produced in plutonium breeding~\cite{Cogswell2016}.

\section*{Acknowledgments}
We thank P. Barbeau and K. Schollberg for useful discussions.  The work
of MB was supported by the National Science Foundation REU grant
number 1757087. The work of PH was supported by the U.S.  Department
of Energy Office of Science under award number DE-SC00018327 and the
National Nuclear Security Administration Office of Defense Nuclear
Nonproliferation R\&D through the consortium for Monitoring,
Technology and Verification under award number DE-NA0003920.

\bibliography{refs,refs2}

\end{document}